\documentclass[english]{revtex4-1}
\usepackage[T1]{fontenc}
\usepackage[latin9]{inputenc}
\usepackage{amsmath}
\usepackage{amssymb}
\usepackage{graphicx}

\makeatletter

\providecommand{\tabularnewline}{\\}

\numberwithin{equation}{section}

\usepackage{babel}

\makeatother

\usepackage{babel}
\begin{document}

\title{Ladder operators and coherent states for multi-step supersymmetric
rational extensions of the truncated oscillator}

\author{Scott E. Hoffmann\textsuperscript{1}, Véronique Hussin\textsuperscript{2},
Ian Marquette\textsuperscript{1} and Yao-Zhong Zhang\textsuperscript{1}}

\address{\textsuperscript{1}School of Mathematics and Physics, The University
of Queensland, Brisbane, QLD 4072, Australia~\\
~\\
\textsuperscript{2}Département de Mathématiques et de Statistique,
Université de Montréal, Montréal, Québec, H3C 3J7, Canada}
\email{scott.hoffmann@uqconnect.edu.au}

\begin{abstract}
We construct ladder operators, $\tilde{C}$ and $\tilde{C}^{\dagger},$
for a multi-step rational extension of the harmonic oscillator on
the half plane, $x\geq0.$ These ladder operators connect all states
of the spectrum in only infinite-dimensional representations of their
polynomial Heisenberg algebra. For comparison, we also construct two
different classes of ladder operator acting on this system that form
finite-dimensional as well as infinite-dimensional representations
of their respective polynomial Heisenberg algebras. For the rational
extension, we construct the position wavefunctions in terms of exceptional
orthogonal polynomials. For a particular choice of parameters, we
construct the coherent states, eigenvectors of $\tilde{C}$ with generally
complex eigenvalues, $z,$ as superpositions of a subset of the energy
eigenvectors. Then we calculate the properties of these coherent states,
looking for classical or non-classical behaviour. We calculate the
energy expectation as a function of $|z|.$ We plot position probability
densities for the coherent states and for the even and odd cat states
formed from these coherent states. We plot the Wigner function for
a particular choice of $z.$ For these coherent states on one arm
of a beamsplitter, we calculate the two excitation number distribution
and the linear entropy of the output state. We plot the standard deviations
in $x$ and $p$ and find no squeezing in the regime considered. By
plotting the Mandel $Q$ parameter for the coherent states as a function
of $|z|,$ we find that the number statistics is sub-Poissonian.
\end{abstract}
\maketitle

\section{Introduction}

Supersymmetric quantum mechanics \cite{Witten1981,Witten1982,Mielnik1984,Adrianov1993,Adrianov1995,Cooper1995,Fernandez1998a,Fernandez1998b,Samsonov1999,Mielnick2000,Carinena2001,Aoyama2001,Carballo2004,Marquette2009,Fernandez2010,Quesne2011,Marquette2012,GomezUllate2014}
has been widely used to create partner Hamiltonians for a given, exactly
solvable, Hamiltonian that have at least part of the spectrum in common
with the original. Typically there are one or more energy levels of
the partner Hamiltonian below the ground state energy of the original
Hamiltonian. 

In this paper, following \cite{Marquette2014b}, we consider multi-step
rational extensions of the harmonic oscillator. The novel aspect of
this paper is that we consider multi-step rational extensions of the
\textit{truncated} oscillator, confined to the half plane $x\geq0,$
with the potential being infinite for $x<0.$ This potential has been
considered by other authors \cite{Fernandez2014,Fernandez2016}, but
not with multi-step rational extensions. Fernández C. \textit{et al.
\cite{Fernandez2018b} }considered the truncated oscillator with the
same rational extension that we use. They constructed coherent states
associated with linearized versions of the ladder operators we label
$\widetilde{L}$ and $\widetilde{L}^{\dagger}$ in Section II below.

In Section II, we review the construction of the partner Hamiltonian
in the untruncated case and the construction of three types of ladder
operator that connect the energy levels. Then we consider the truncated
case, which eliminates all states with position wavefunctions even
in $x.$ We construct three types of ladder operator that connect
the energy levels of the truncated system. We will find, below, that
there is significant structure to the energy levels, such that some
of the levels are singlets or doublets under the action of the various
ladder operators. In every case there are also infinite dimensional
representations of the polynomial Heisenberg algebras.

We note that the potential we are considering is singular, taking
an infinite value for all $x<0.$ Other work has been done on potentials
with singularities \cite{Marquez1998}.

In Section III we construct coherent states associated with the ladder
operators we label $(\tilde{C},\tilde{C}^{\dagger}).$ Coherent states
are of widespread interest \cite{Glauber1963a,Glauber1963b,Klauder1963a,Klauder1963b,Barut1971,Perelomov1986,Gazeau1999,Quesne2001,Fernandez1995,Fernandez1999,GomezUllate2014,Ali2014}
because they can, in some cases, be the quantum-mechanical states
with the most classical behaviour. Then we systematically investigate
properties of these coherent states, looking for classical or non-classical
behaviour. We calculate the energy expectation as a function of $|z|,$
where $z$ is the generally complex coherent state parameter. We calculate
position probability densities for a particular choice of parameters.
A similar calculation is done for even and odd cat states. We plot
the Wigner function, an accepted measure of classicality, for a particular
choice of parameters. For our coherent state on one arm of a beamsplitter,
we calculate the two excitation (``photon'') number distribution
and look for factorizability. We calculate the linear entropy of the
output state as a measure of entanglement. We calculate the standard
deviations in position and momentum over a range of $z$ values to
look for squeezing and to confirm the Heisenberg uncertainty principle.
Lastly, we calculate the Mandel $Q$ parameter for these coherent
states as a function of $|z|$ to characterize the number statistics.

In Section IV, we compare our results with those of Fernández C. \textit{et
al. }\cite{Fernandez2018b}\textit{.}

Conclusions follow in Section IV.

\section{Theory of multi-step rational extensions of the harmonic oscillator}

\subsection{The untruncated oscillator}

We scale the harmonic oscillator Hamiltonian,
\begin{equation}
H_{HO}=-\frac{\hbar^{2}}{2m}\frac{d^{2}}{dx^{2}}+\frac{1}{2}m\omega^{2}x^{2},\label{eq:101}
\end{equation}
according to $\hbar=1,$ $m=1/2,$ $\omega=2,$ to give the (1) Hamiltonian
\begin{equation}
H^{(1)}=-\frac{d^{2}}{dx^{2}}+x^{2},\label{eq:102}
\end{equation}
with energy levels $E_{\nu}^{(1)}=2\nu+1$ for $\nu=0,1,2,\dots.$
The annihilation and creation operators are, respectively,
\begin{equation}
a=\frac{d}{dx}+x,\quad a^{\dagger}=-\frac{d}{dx}+x\label{eq:102.1}
\end{equation}
satisfying the Heisenberg algebra
\begin{equation}
[H^{(1)},a^{\dagger}]=2a^{\dagger},\quad[H^{(1)},a]=-2a,\quad[a,a^{\dagger}]=2.\label{eq:102.2}
\end{equation}

We first present the $k$-step construction for the untruncated oscillator,
for $x$ on the entire real axis, following the results of \cite{Marquette2014b}
on Darboux-Crum (state adding) and Krein-Adler (state deleting) SUSY
QM. The class of example chains we will consider in this paper is
given by consecutive values of $m:$
\begin{equation}
\{m_{1},m_{2},\dots,m_{k-1},m_{k}\}=\{2,3,\dots,k,k+1\},\label{eq:2.1}
\end{equation}
with $k$ even.

We choose for the state adding case the following seed solutions
\begin{equation}
\phi_{m_{1}}(x)=\mathcal{H}_{m_{1}}(x)\,e^{x^{2}/2},\dots,\phi_{m_{k}}(x)=\mathcal{H}_{m_{k}}(x)\,e^{x^{2}/2},\label{eq:2.2}
\end{equation}
where
\begin{equation}
\mathcal{H}_{m}(x)=(-i)^{m}H_{m}(ix)\label{eq:103}
\end{equation}
are the modified Hermite polynomials in terms of the standard Hermite
polynomials $H_{m}(x)$ \cite{Gradsteyn1980}. First we define
\begin{align}
Q^{(1)} & =\phi_{m_{1}},\nonumber \\
Q^{(i)} & =\frac{\mathcal{W}(\phi_{m_{1}},\dots,\phi_{m_{i}})}{\mathcal{W}(\phi_{m_{1}},\dots,\phi_{m_{i-1}})}\quad\mathrm{for}\ i=2,\dots,k,\label{eq:2.3}
\end{align}
where $\mathcal{W}$ indicates the Wronskian determinant
\begin{equation}
\mathcal{W}(\phi_{1},\phi_{2},\dots,\phi_{n})=\det\begin{pmatrix}\phi_{1} & \phi_{2} & \dots & \phi_{n}\\
\phi_{1}^{(1)} & \phi_{2}^{(1)} & \dots & \phi_{n}^{(1)}\\
\dots & \dots & \dots & \dots\\
\phi_{1}^{(n-1)} & \phi_{2}^{(n-1)} & \dots & \phi_{n}^{(n-1)}
\end{pmatrix}.\label{eq:104}
\end{equation}
The supercharges, $A^{(i)},$ are given by
\begin{align}
A^{(i)} & =\frac{d}{dx}+W^{(i)}\quad\mathrm{for}\ i=1,\dots,k,\label{eq:2.4}
\end{align}
with
\begin{equation}
W^{(i)}=-\frac{d}{dx}\ln Q^{(i)}\quad\mathrm{for}\ i=1,\dots,k.\label{eq:2.5}
\end{equation}

The product of these supercharges is
\begin{align}
A & =A^{(k)}A^{(k-1)}\dots A^{(1)}.\label{eq:2.6}
\end{align}
It is not the case here that $H^{(1)}=A^{\dagger}A,H^{(2)}=AA^{\dagger},$
as in simpler examples of supersymmetry with first-order supercharges.
However we require that the intertwining relation,
\begin{equation}
A\,H^{(1)}=H^{(2)}\,A,\label{eq:2.8-1}
\end{equation}
hold and relate the initial Hamiltonian $H^{(1)}$ (the harmonic oscillator)
to a final Hamiltonian $H^{(2)}$ through a chain of Hamiltonians,
alternately regular and singular. We solve this relation to give
\begin{align}
H^{(1)} & =-\frac{d^{2}}{dx^{2}}+x^{2},\quad H^{(2)}=-\frac{d^{2}}{dx^{2}}+V^{(2)},\nonumber \\
V^{(2)} & =x^{2}-2\frac{d^{2}}{dx^{2}}\ln\mathcal{W}(\phi_{m_{1}},\dots,\phi_{m_{k}})+\Delta E.\label{eq:2.7}
\end{align}

We know the wavefunctions of $H^{(1)}$
\begin{equation}
\psi_{\nu}=H_{\nu}\,e^{-x^{2}/2},\label{eq:2.9}
\end{equation}
with energies $E_{\nu}^{(1)}=2\nu+1,\ \mathrm{for}\ \nu=0,1,2,\dots.$
The partner Hamiltonian wavefunctions can be written in terms of the
seed solutions
\begin{align}
\psi_{\nu}^{(2)} & \propto\mathcal{W}(\phi_{m_{1}},\dots,\phi_{m_{k}},\psi_{\nu})\quad\mathrm{for}\ \nu=0,1,2,\dots,\nonumber \\
\psi_{-m_{i}-1}^{(2)} & \propto\mathcal{W}(\phi_{m_{1}},\dots,\check{\phi}_{m_{i}},\dots,\phi_{m_{k}})\quad\mathrm{for}\ i=1,2,\dots,k,\label{eq:2.10}
\end{align}
where $\mathcal{W}(\phi_{m_{1}},\dots,\check{\phi}_{m_{i}},\dots,\phi_{m_{k}})$
means $\phi_{m_{i}}$ is missing from the sequence. Here $\psi_{\nu}^{(2)}$
are the states obtained from $\psi_{\nu}^{(1)}$ by $A\psi_{\nu}^{(1)}.$
The other states, $\psi_{-m_{i}-1}^{(2)},$ are obtained via the constraint
$A^{\dagger}\psi=0.$ Here $A^{\dagger}$ is a $k-$th order differential
operator and there is more than one zero mode.

The energy levels are $E_{\nu}^{(2)}=2\nu+1$ for $\nu=-m_{k}-1,-m_{k-1}-1,\dots,-m_{1}-1,0,1,2,\dots.$

The other construction (state deleting) uses only the seed solutions
$\{\psi_{k},\psi_{k+1}\},$ with the other states having been deleted.

We construct the supercharges and superpotentials in a similar way
to what was done in the first construction, starting with
\begin{align}
\bar{Q}^{(k)} & =\psi_{k},\nonumber \\
\bar{Q}^{(k+1)} & =\frac{\mathcal{W}(\psi_{k},\psi_{k+1})}{\psi_{k}}.\label{eq:2.12}
\end{align}
Then
\begin{equation}
\overline{W}^{(i)}=-\frac{d}{dx}\ln\overline{Q}^{(i)}\quad\mathrm{for}\ i=k,k+1,\label{eq:105}
\end{equation}
\begin{align}
\overline{A}^{(i)} & =\frac{d}{dx}+\overline{W}^{(i)}\quad\mathrm{for}\ i=k,k+1,\label{eq:106}
\end{align}
and
\begin{equation}
\overline{A}=\overline{A}^{(k+1)}\overline{A}^{(k)}.\label{eq:107}
\end{equation}

The partner Hamiltonian is again equal to $H^{(2)}$ as in Eq.~(\ref{eq:2.7}),
found by solving the intertwining relation
\begin{equation}
\overline{A}\,H^{(1)}=H^{(2)}\,\overline{A}\label{eq:2.8-1-1}
\end{equation}
with an appropriate energy shift. The wavefunctions and energy levels
are then the same as in the previous construction \cite{Marquette2014b}. 

Two types of standard ladder operators can be generated, $(L^{\dagger},L)$
and $(\bar{L}^{\dagger},\bar{L}),$ given by
\begin{align}
L^{\dagger} & =Aa^{\dagger}A^{\dagger},\quad L=AaA^{\dagger},\label{eq:2.13}
\end{align}
and
\begin{align}
\bar{L}^{\dagger} & =\bar{A}a^{\dagger}\bar{A}^{\dagger},\quad\bar{L}=\bar{A}a\bar{A}^{\dagger}.\label{eq:2.14}
\end{align}

We find that the commutators for $(L^{\dagger},L)$ are
\begin{align}
[H^{(2)},L^{\dagger}] & =2L^{\dagger},\quad[H^{(2)},L]=-2L,\quad[L,L^{\dagger}]=P(H^{(2)}+2)-P(H^{(2)}),\label{eq:2.15}
\end{align}
with
\begin{equation}
P(H^{(2)})=\prod_{i=1}^{k}(H^{(2)}+2m_{i}-1)(H^{(2)}-1)\prod_{j=1}^{k}(H^{(2)}+2m_{j}+1).\label{eq:2.16}
\end{equation}

We find that the commutators for $(\bar{L}^{\dagger},\bar{L})$ are
\begin{align}
[H^{(2)},\bar{L}^{\dagger}] & =2\bar{L}^{\dagger},\quad[H^{(2)},\bar{L}]=-2\bar{L},\quad[\bar{L},\bar{L}^{\dagger}]=\bar{P}(H^{(2)}+2)-\bar{P}(H^{(2)}),\label{eq:2.17}
\end{align}
with
\begin{equation}
\bar{P}(H^{(2)})=\prod_{i=1}^{m_{k}}(H^{(2)}+2m_{k}-2i-1)(H^{(2)}+2m_{k}+1)\prod_{j=1}^{m_{k}}(H^{(2)}+2m_{k}-2j+1),\label{eq:2.18}
\end{equation}
for $j\neq m_{k}-m_{k-1},\dots,m_{k}-m_{1}.$

The combination of both Darboux-Crum and Krein-Adler paths can be
exploited to generate a different type of ladder operator, the type
we will focus on in this paper,
\begin{align}
C & =\bar{A}A^{\dagger},\quad C^{\dagger}=A\bar{A}^{\dagger}.\label{eq:2.19}
\end{align}
We find that the commutators are
\begin{align}
[H^{(2)},C^{\dagger}] & =(2m_{k}+2)C^{\dagger},\quad[H^{(2)},C]=-(2m_{k}+2)C,\quad[C,C^{\dagger}]=Q(H^{(2)}+2m_{k}+2)-Q(H^{(2)}),\label{eq:2.20}
\end{align}
with
\begin{equation}
Q(H^{(2)})=\prod_{i=1}^{k}(H^{(2)}+2m_{i}+1)\prod_{j\neq m_{k}-m_{k-1},\dots,m_{k}-m_{1}}^{m_{k}}(H^{(2)}-2j-1).\label{eq:2.21}
\end{equation}
The zero modes of $(C,C^{\dagger}),(L,L^{\dagger})$ and $(\bar{L},\bar{L}^{\dagger})$
are associated with the energies for which $Q,P$ and $\bar{P}$ and
$Q(H^{(2)}+2m_{k}+2),P(H^{(2)}+2)$ and $\bar{P}(H^{(2)}+2)$ vanish.

In this paper we will consider the untruncated and truncated oscillators
and their supersymmetric partners for the particular choice
\begin{equation}
\{m_{1},m_{2},m_{3},m_{4}\}=\{2,3,4,5\}.\label{eq:2.22}
\end{equation}
We find that the partner potential is
\begin{align}
V^{(2)} & =-8+x^{2}-\frac{1024(-315+90x^{2}-1020x^{4}+328x^{6})}{(45+120x^{4}-64x^{6}+16x^{8})^{2}}+\nonumber \\
 & \quad+\frac{64(-112-13x^{2}-4x^{4}+4x^{6})}{45+120x^{4}-64x^{6}+16x^{8}}.\label{eq:2.23}
\end{align}
This is in full agreement with \cite{Fernandez2018}, after noting
that their Hamiltonians are one half of ours.

The first (state adding) path uses
\begin{align}
Q^{(1)} & =e^{x^{2}/2}(2+4x^{2}),\nonumber \\
Q^{(2)} & =8e^{x^{2}/2}\frac{3+4x^{2}}{2+4x^{2}},\nonumber \\
Q^{(3)} & =16e^{x^{2}/2}\frac{9+18x^{2}-12x^{4}+8x^{6}}{3+4x^{4}},\nonumber \\
Q^{(4)} & =96e^{x^{2}/2}\frac{9+18x^{2}(15-8x^{2}+2x^{4})}{9+18x^{2}-12x^{4}+8x^{6}}.\label{eq:2.24}
\end{align}

The second (state deleting) path uses
\begin{align}
\bar{Q}^{(4)} & =e^{-x^{2}/2}(12-48x^{2}+16x^{4}),\nonumber \\
\bar{Q}^{(5)} & =8e^{-x^{2}/2}\frac{(45+8x^{4}(15-8x^{2}+2x^{4}))}{3-12x^{2}+4x^{4}}.\label{eq:2.25}
\end{align}

We can form the $C,L$ and $\bar{L}$ ladder operators from
\begin{align}
C & =\bar{A}^{(5)}\bar{A}^{(4)}A^{(1)\dagger}A^{(2)\dagger}A^{(3)\dagger}A^{(4)\dagger},\nonumber \\
L & =A^{(4)}A^{(3)}A^{(2)}A^{(1)}aA^{(1)\dagger}A^{(2)\dagger}A^{(3)\dagger}A^{(4)\dagger},\nonumber \\
\bar{L} & =\bar{A}^{(5)}\bar{A}^{(4)}a\bar{A}^{(4)\dagger}\bar{A}^{(5)\dagger}.\label{eq:2.26}
\end{align}
Then $C^{\dagger},L^{\dagger}$ and $\bar{L^{\dagger}}$ are just
the respective Hermitian conjugates. For $\{l_{1},l_{2},l_{3}\}=\{C,L,\bar{L}\},$
the polynomial Heisenberg algebras that they satisfy take the forms
for our particular model
\begin{equation}
[H^{(2)},l_{n}]=-\lambda_{n}l_{n},\quad[H^{(2)},l_{n}^{\dagger}]=+\lambda_{n}l_{n}^{\dagger},\quad[l_{n},l_{n}^{\dagger}]=P_{n}(H^{(2)}+\lambda_{n})-P_{n}(H^{(2)}),\label{eq:2.27}
\end{equation}
with
\begin{equation}
\{\lambda_{1},\lambda_{2},\lambda_{3}\}=\{2,2,12\},\label{eq:2.28}
\end{equation}
and
\begin{align}
P_{1}(H^{(2)}) & =(H^{(2)}-11)(H^{(2)}-9)(H^{(2)}+5)(H^{(2)}+7)(H^{(2)}+9)(H^{(2)}+11),\nonumber \\
P_{2}(H^{(2)}) & =(H^{(2)}+3)(H^{(2)}+5)(H^{(2)}+7)(H^{(2)}+9)(H^{(2)}-1)(H^{(2)}+5)(H^{(2)}+7)(H^{(2)}+9)(H^{(2)}+11),\nonumber \\
P_{3}(H^{(2)}) & =(H^{(2)}+11)(H^{(2)}+1)(H^{(2)}-1)(H^{(2)}+3)(H^{(2)}+1).\label{eq:2.29}
\end{align}

The zero modes of the ladder operators are given in Table 1.

\begin{table}
\begin{centering}
\begin{tabular}{|c|c|}
\hline 
Ladder operator & Zero modes\tabularnewline
\hline 
\hline 
$C$ & $\psi_{-6}^{(2)},\psi_{-5}^{(2)},\psi_{-4}^{(2)},\psi_{-3}^{(2)},\psi_{4}^{(2)},\psi_{5}^{(2)}$\tabularnewline
\hline 
$C^{\dagger}$ & No physical states\tabularnewline
\hline 
$L$ & $\psi_{-6}^{(2)},\psi_{-5}^{(2)},\psi_{-4}^{(2)},\psi_{-3}^{(2)},\psi_{0}^{(2)}$\tabularnewline
\hline 
$L^{\dagger}$ & $\psi_{-6}^{(2)},\psi_{-5}^{(2)},\psi_{-4}^{(2)},\psi_{-3}^{(2)}$\tabularnewline
\hline 
$\bar{L}$ & $\psi_{-6}^{(2)},\psi_{0}^{(2)}$\tabularnewline
\hline 
$\bar{L}^{\dagger}$ & $\psi_{-3}^{(2)}$\tabularnewline
\hline 
\end{tabular}
\par\end{centering}
\caption{Zero modes of the six ladder operators for the untruncated system.}
\end{table}

The energy levels of $H^{(2)}$ are given by $E_{\nu}^{(2)}=2\nu+1\quad\mathrm{for}\ \nu=-6,-5,-4,-3,0,1,2,\dots.$
(Note we do not apply an energy shift here to make all energies positive).

In Figure 1 we show the physical states and the actions of the three
types of ladder operator for the untruncated case we are considering.
We note that $(L,L^{\dagger})$ and $(\bar{L},\bar{L}^{\dagger})$
have finite-dimensional as well as infinite-dimension representations
of their respective polynomial Heisenberg algebras, while $(C,C^{\dagger})$
have only infinite-dimensional representations.

\begin{figure}
\begin{centering}
\includegraphics[width=10cm]{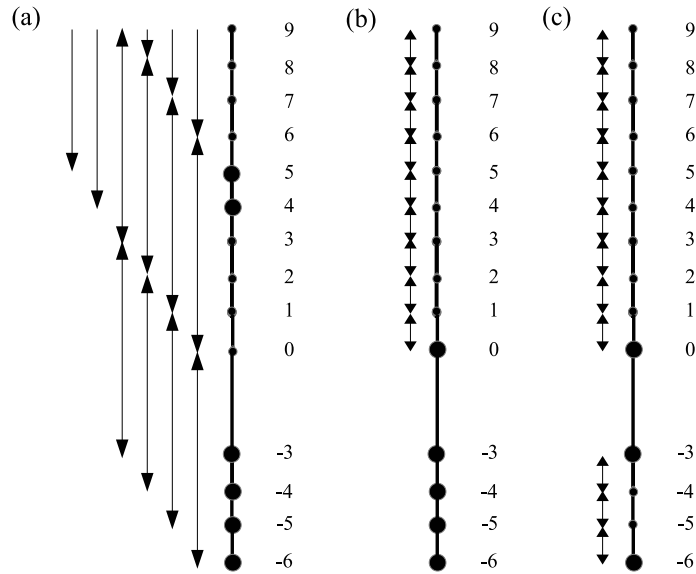}
\par\end{centering}
\caption{Physical states and actions of the ladder operators (a) $C,C^{\dagger},$
(b) $L,L^{\dagger}$ and (c) $\bar{L},\bar{L}^{\dagger}.$ The large
dots represent zero modes of either raising or lowering operators
in each case.}

\end{figure}

\subsection{The truncated oscillator}

The truncated oscillator has Hamiltonian
\begin{equation}
\tilde{H}^{(1)}=\begin{cases}
-\frac{d^{2}}{dx^{2}}+x^{2} & 0\leq x<\infty,\\
\infty & x<0.
\end{cases}\label{eq:2.34}
\end{equation}
In this case only the odd states from the previous derivation (Section
II. A) are the physical states with wavefunctions satisfying the boundary
condition of vanishing at the origin. To obtain the truncated supersymmetric
partner we also select the physical states which satisfy the boundary
condition. SUSY QM might only be formal if the boundary condition
of the Hamiltonian and superpartner are not the same.

The physical state wavefunctions of $\tilde{H}^{(1)}$ are
\begin{align}
\tilde{\psi}_{1}^{(1)} & =2x\,e^{-x^{2}/2},\nonumber \\
\tilde{\psi}_{3}^{(1)} & =(-12x+8x^{3})\,e^{-x^{2}/2},\nonumber \\
\tilde{\psi}_{5}^{(1)} & =(120x-160x^{3}+32x^{5})\,e^{-x^{2}/2},\label{eq:2.35}\\
 & \dots\nonumber 
\end{align}
with energies $\tilde{E}_{\nu}^{(1)}=2\nu+1,\ \mathrm{for}\ \nu=1,3,5,\dots.$
For the superpartner, they are
\begin{align}
\psi_{-5}^{(2)},\quad\tilde{E}_{-5}^{(2)} & =-9,\nonumber \\
\psi_{-3}^{(2)},\quad\tilde{E}_{-3}^{(2)} & =-5,\nonumber \\
\psi_{1}^{(2)},\quad E_{1}^{(2)} & =3,\nonumber \\
\psi_{3}^{(2)},\quad\tilde{E}_{3}^{(2)} & =7,\nonumber \\
 & \dots\label{eq:2.36}
\end{align}
with the $\psi_{\nu}^{(2)}$ wavefunctions given by Eq. (\ref{eq:2.10}).

The ladder operators $C^{\dagger}$ and $C$ change the index by $\pm6,$
respectively, an even number, so they transform odd states into odd
states and even states into even states. Thus they split the state
space into two distinct, irreducible representations. Hence the ladder
operators $\tilde{C}$ and $\tilde{C}^{\dagger}$ for the truncated
oscillator can be taken as $C$ and $C^{\dagger},$ respectively,
restricted to the odd subspace.

The ladder operators $L^{\dagger},L,\bar{L}^{\dagger}$ and $\bar{L}$
change the index by $\pm1,$ mixing odd states and even states. Thus
we construct new lowering operators for the truncated oscillator that
change the index by $-2$:
\begin{align}
\widetilde{L} & =Aa^{2}A^{\dagger},\nonumber \\
\widetilde{\bar{L}} & =\bar{A}a^{2}\bar{A}^{\dagger},\label{eq:2.37}
\end{align}
with the raising operators given by the Hermitian conjugates, which
change the index by $+2$. The polynomial Heisenberg algebras take
the forms
\begin{align}
[\widetilde{H}^{(2)},\widetilde{L}] & =-4\widetilde{L},\quad[\widetilde{H}^{(2)},\widetilde{L}^{\dagger}]=4\widetilde{L}^{\dagger},\nonumber \\{}
[\widetilde{L},\widetilde{L}^{\dagger}] & =\widetilde{P(}\widetilde{H}^{(2)}+4)-\widetilde{P(}\widetilde{H}^{(2)})\label{eq:2.38}
\end{align}
and
\begin{align}
[\widetilde{H}^{(2)},\widetilde{\bar{L}}] & =-4\widetilde{\bar{L}},\quad[\widetilde{H}^{(2)},\widetilde{\bar{L}}^{\dagger}]=4\widetilde{\bar{L}}^{\dagger},\nonumber \\{}
[\widetilde{\bar{L}},\widetilde{\bar{L}}^{\dagger}] & =\widetilde{\bar{P}(}\widetilde{H}^{(2)}+4)-\widetilde{\bar{P}(}\widetilde{H}^{(2)}).\label{eq:2.39}
\end{align}

In these expressions
\begin{align}
\widetilde{P(}\widetilde{H}^{(2)}) & =(\widetilde{H}^{(2)}+1)(\widetilde{H}^{(2)}+3)(\widetilde{H}^{(2)}+5)(\widetilde{H}^{(2)}+7)(\widetilde{H}^{(2)}-5)\times\nonumber \\
 & \quad\times(\widetilde{H}^{(2)}-3)(\widetilde{H}^{(2)}+5)(\widetilde{H}^{(2)}+7)(\widetilde{H}^{(2)}+9)(\widetilde{H}^{(2)}+11),\nonumber \\
\widetilde{\bar{P}(}\widetilde{H}^{(2)}) & =(\widetilde{H}^{(2)}+9)(\widetilde{H}^{(2)}-1)(\widetilde{H}^{(2)}+11)(\widetilde{H}^{(2)}-3)(\widetilde{H}^{(2)}+3)(\widetilde{H}^{(2)}+1).\label{eq:2.40}
\end{align}

The zero modes are shown in Table 2.

\begin{table}
\begin{centering}
\begin{tabular}{|c|c|}
\hline 
Ladder operator & Zero modes\tabularnewline
\hline 
\hline 
$\tilde{C}$ & $\psi_{-5}^{(2)},\psi_{-3}^{(2)},\psi_{5}^{(2)}$\tabularnewline
\hline 
$\tilde{C}^{\dagger}$ & No physical states\tabularnewline
\hline 
$\tilde{L}$ & $\psi_{-5}^{(2)},\psi_{-3}^{(2)},\psi_{1}^{(2)}$\tabularnewline
\hline 
$\tilde{L}^{\dagger}$ & $\psi_{-5}^{(2)},\psi_{-3}^{(2)}$\tabularnewline
\hline 
$\widetilde{\bar{L}}$ & $\psi_{-5}^{(2)},\psi_{1}^{(2)}$\tabularnewline
\hline 
$\widetilde{\bar{L}}^{\dagger}$ & $\psi_{-3}^{(2)}$\tabularnewline
\hline 
\end{tabular}
\par\end{centering}
\caption{Zero modes of four ladder operators for the truncated system.}

\end{table}

In Figure 2 we show the physical states and the actions of the three
types of ladder operator for the truncated case we are considering.
We see that only the ladder operators $(\tilde{C},\tilde{C}^{\dagger})$
have only infinite-dimensional representations.

\begin{figure}
\begin{centering}
\includegraphics[width=10cm]{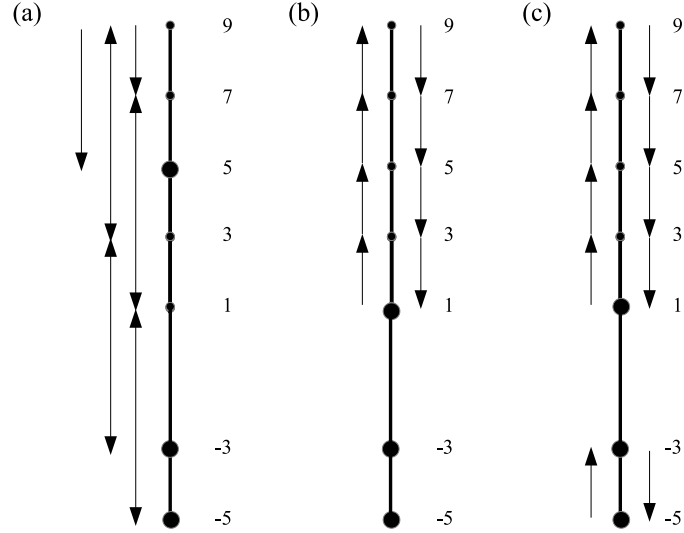}
\par\end{centering}
\caption{Physical states and actions of the ladder operators (a) $\tilde{C},\tilde{C}^{\dagger},$
(b) $\tilde{L},\tilde{L}^{\dagger}$ and (c) $\widetilde{\bar{L}},\widetilde{\bar{L}}^{\dagger}.$
The large dots represent zero modes of either raising or lowering
operators in each case.}

\end{figure}

\section{Construction and properties of coherent states}

We construct the coherent states associated with the ladder operators
$(\tilde{C},\tilde{C}^{\dagger})$ for the particular case of lowest
weight $\mu=5.$

For $\mu=5,$ the matrix elements of $\tilde{C},$ as given in \cite{Marquette2014b},
are
\begin{equation}
\langle\,5+6i-6\,|\,\tilde{C}\,|\,5+6i\,\rangle=a_{5+6i}=[2^{6}(11+6i)\frac{(4+6i)!}{(6i-1)!}\frac{8+6i}{2+6i}\frac{9+6i}{3+6i}\frac{10+6i}{4+6i}]^{\frac{1}{2}},\quad\mathrm{for}\ i=1,2,3,\dots.\label{eq:3.1}
\end{equation}

Using similar results from \cite{Hoffmann2018c}, the coherent states
with the Barut-Girardello definition \cite{Barut1971} as eigenvectors
of the lowering operator with generally complex eigenvalues,
\begin{equation}
\tilde{C}\,|\,z;c,5\,\rangle=z\,|\,z;c,5\,\rangle,\label{eq:3.1.5}
\end{equation}
are given by
\begin{equation}
|\,z;c,5\,\rangle=\sum_{k=0}^{\infty}|\,5+6k\,\rangle\alpha_{k}^{(5)}\label{eq:3.2}
\end{equation}
with
\begin{equation}
\alpha_{k}^{(5)}(z)=\frac{1}{\sqrt{F^{(5)}(z)}}\frac{z^{k}}{D_{k}^{(5)}}.\label{eq:3.3}
\end{equation}
Here
\begin{equation}
D_{0}^{(5)}=1,\quad D_{k}^{(5)}=\prod_{i=1}^{k}a_{5+6i}\quad\mathrm{for}\ k\geq1\label{eq:3.4}
\end{equation}
and
\begin{equation}
F^{(5)}(z)=\sum_{k=0}^{\infty}\frac{|z|^{2k}}{D_{k}^{(5)2}}.\label{eq:3.5}
\end{equation}

\subsection{Energy expectation}

We first calculate the energy expectation as a function of $|z|$
for this coherent state. This is
\begin{equation}
\langle\,E^{(5)}(|z|)\,\rangle=\langle\,z;c,5\,|\,H^{(2)}\,|\,z;c,5\,\rangle=11+\frac{12}{F^{(5)}(z)}\sum_{k=0}^{\infty}k\frac{|z|^{2k}}{D_{k}^{(5)2}}.\label{eq:3.6}
\end{equation}
The denominators grow very rapidly: $D_{k}^{(5)2}=1,1.5\times10^{8},3.3\times10^{17},4.1\times10^{27}$
for $k=0,1,2,3.$ We take the sum to $k=6$ with negligible remainder.
As a consequence, $F^{(5)}(z)$ and $\langle\,E^{(5)}(z)\,\rangle$
grow very slowly for low values of $|z|.$ We find the following profile
up to $|z|=10^{5},$ as shown in Fig. 3. Fernández C. \textit{et al.}
\cite{Fernandez2018b} did not calculate the energy expectation.
\begin{figure}
\noindent \begin{centering}
\includegraphics[width=8cm]{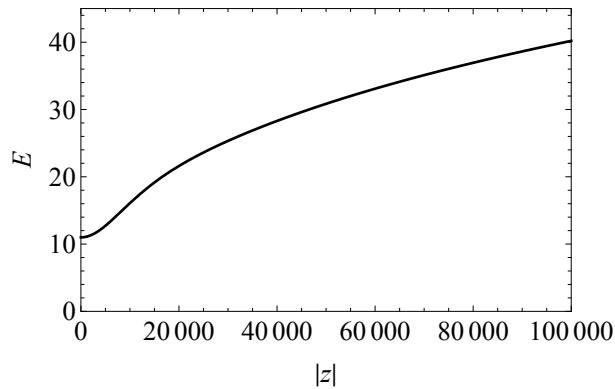}
\par\end{centering}
\caption{Energy expectation for the coherent states with $\mu=5.$}
\end{figure}

\subsection{Position probability densities}

According to Marquette and Quesne \cite{Marquette2014b}, the unnormalized
wavefunctions are
\begin{equation}
\Psi_{\nu}^{(2)}(x)=\frac{\mathcal{W}(\phi_{2},\phi_{3},\phi_{4},\phi_{5},\psi_{\nu})}{\mathcal{W}(\phi_{2},\phi_{3},\phi_{4},\phi_{5})},\quad\mathrm{where}\ \psi_{\nu}(x)=H_{\nu}(x)\,e^{-x^{2}/2}.\label{eq:3.7}
\end{equation}
We need to calculate
\begin{equation}
\mathcal{N}_{\nu}=\int_{0}^{\infty}dx\,|\Psi_{\nu}^{(-)}(x)|^{2}\label{eq:3.8}
\end{equation}
(for normalization on the half line) and then form the normalized
wavefunctions
\begin{equation}
\psi_{\nu}^{(-)}(x)=\frac{\Psi_{\nu}^{(-)}(x)}{\sqrt{\mathcal{N}_{\nu}}}.\label{eq:3.9}
\end{equation}

Then the probability density for the coherent state is
\begin{equation}
\rho(x,t;z,5)=|\langle\,x\,|\,e^{-iH^{(2)}t}\,|\,z;c,5\,\rangle|^{2}=\left|\sum_{k=0}^{\infty}\psi_{5+6k}^{(-)}(x)\frac{1}{\sqrt{F^{(5)}(z)}}\frac{(z\,e^{-i12t})^{k}}{D_{k}^{(5)}}\right|^{2}.\label{eq:3.10}
\end{equation}
For $|z|\leq10^{5},$ we can take the sum to $k=11.$ The density
plot over one period is shown in Figure 4. We see considerable structure,
with individual wavepackets diminishing and reforming.
\begin{figure}
\noindent \begin{centering}
\includegraphics[width=6cm]{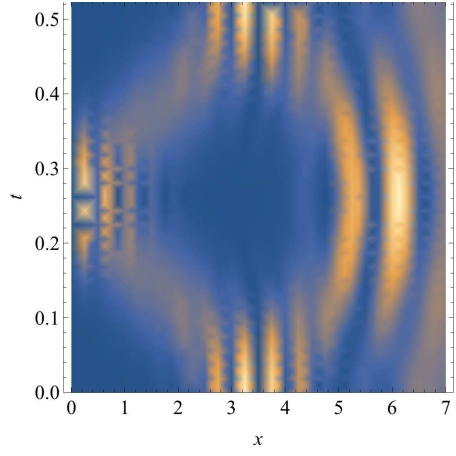}
\par\end{centering}
\caption{Density plot of $\rho(x,t;10^{5},5)$ over one period.}
\end{figure}

\subsection{Even and odd cat states}

The distinguishability measure in this case is
\begin{align}
D^{(5)}(z) & =\langle\,-z;c,5\,|\,+z;c,5\,\rangle=\frac{1}{F^{(5)}(z)}\sum_{k=0}^{\infty}(-)^{k}\frac{|z|^{2k}}{D_{k}^{(5)2}}.\label{eq:3.11}
\end{align}
This function is plotted in Figure 5. We see that the distinguishability
falls to zero as $|z|\rightarrow\infty,$ but there are also two other
zeros at finite values of $|z|.$
\begin{figure}
\noindent \begin{centering}
\includegraphics[width=8cm]{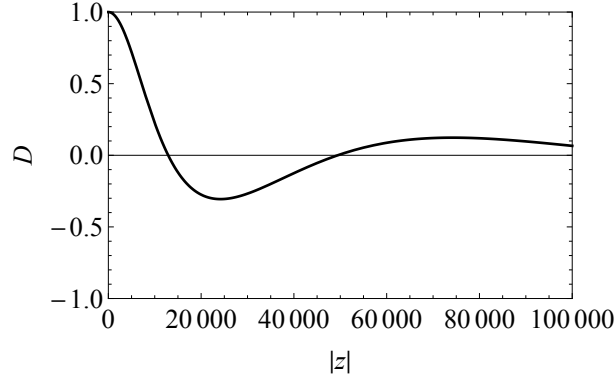}
\par\end{centering}
\caption{Distinguishability measure $D^{(5)}(|z|).$}
\end{figure}

We form the cat states (approximately normalized)
\begin{align}
|\,z;+,c,5\,\rangle & =\frac{1}{\sqrt{2}}\{|\,+z;c,5\,\rangle+|\,-z;c,5\,\rangle\},\quad|\,z;-,c,5\,\rangle=\frac{1}{\sqrt{2}}\{|\,+z;c,5\,\rangle-|\,-z;c,5\,\rangle\},\label{eq:3.12}
\end{align}
and plot their time-dependent probability densities for $z=10^{5}$
(to ensure distinguishability) in Figure 6. (Again, we take the sum
over $k$ to $k=11.$) Here, too, we see a great deal of structure.
\begin{figure}
\noindent \begin{centering}
\includegraphics[width=12cm]{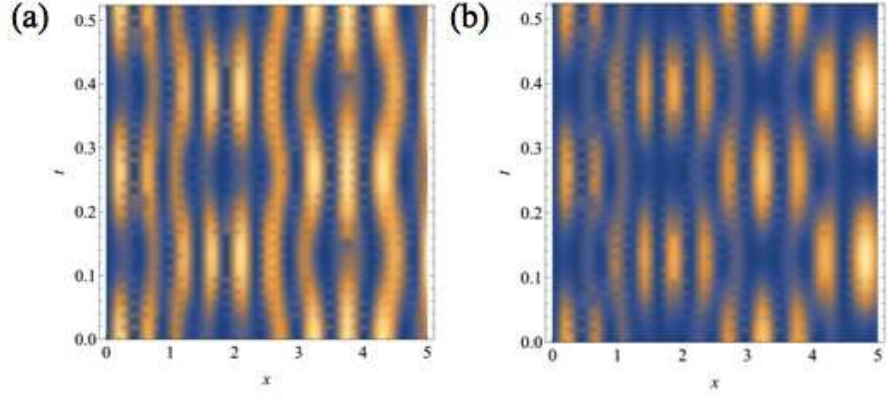}
\par\end{centering}
\caption{Time-dependent densities for (a) the + and (b) the - cat states, with
$z=10^{5}$.}
\end{figure}

\subsection{Wigner function}

We use the formula for the Wigner function \cite{Hoffmann2018c}
\begin{equation}
W(x,p;z,5)=\sum_{k_{1}=0}^{\infty}\sum_{k_{2}=0}^{\infty}\alpha_{k_{1}}^{(5)*}(z)w_{k_{1}k_{2}}(x,p)\alpha_{k_{2}}^{(5)}(z)\label{eq:3.13}
\end{equation}
with
\begin{equation}
w_{k_{1}k_{2}}(x,p)=\begin{cases}
\frac{1}{\pi}\int_{-x}^{x}dy\,\psi_{5+6k_{1}}^{(-)*}(x-y)\psi_{5+6k_{2}}^{(-)}(x+y)e^{-i2py} & x>0,\\
0 & x<0
\end{cases}\label{eq:3.14}
\end{equation}
in this truncated case. The integrand can only be nonzero for $x-y>0$
and $x+y>0.$ This gives $-x<y<x$ for positive $x$ and no region
of nonzero integrand for negative $x.$

For $z=500,$ we assume the $k_{1}=k_{2}=0$ contribution is sufficient
because of the rapid decrease with $k$ of the coefficients $\alpha_{k}^{(5)}(500)$.
We find the Wigner function displayed in Figure 7, vanishing for negative
$x$. Closer inspection shows only small areas of small negative Wigner
function $(W(1.9,0.8;500,5)=-0.036)$, which may disappear if we improve
our approximation. We conclude that there is no clear signature of
non-classical behaviour.
\begin{figure}
\noindent \begin{centering}
\includegraphics[width=8cm]{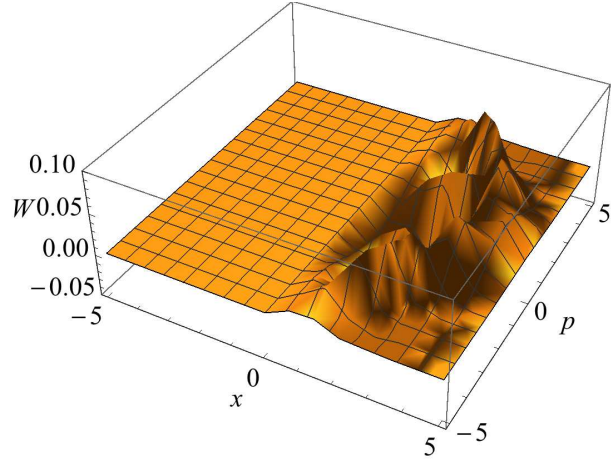}
\par\end{centering}
\caption{Wigner function for $z=500$.}
\end{figure}

\subsection{Coherent state on a beamsplitter}

For our coherent state placed on one arm of a beamsplitter, the probability
of measuring $n_{1}$ ``photons'' in the first output arm and $n_{2}$
``photons'' in the second output arm is \cite{Hoffmann2018c}
\begin{align}
P^{(5)}(n_{1},n_{2};z) & =\left|\alpha_{n_{1}+n_{2}}^{(5)}(z)\frac{1}{2^{\frac{n_{1}+n_{2}}{2}}}\binom{n_{1}+n_{2}}{n_{2}}^{\frac{1}{2}}i^{n_{2}}\right|^{2}\nonumber \\
 & =\frac{(n_{1}+n_{2})!}{n_{1}!n_{2}!}\frac{1}{F^{(5)}(z)}\frac{1}{|D_{n_{1}+n_{2}}^{(5)}|^{2}}(\frac{|z|^{2}}{2})^{n_{1}+n_{2}}.\label{eq:3.15}
\end{align}
Clearly this does not factorize, as seen by the presence of functions
of $n_{1}+n_{2}$. We plot this in Figure 8 compared to the harmonic
oscillator case.
\begin{figure}
\noindent \begin{centering}
\includegraphics[width=12cm]{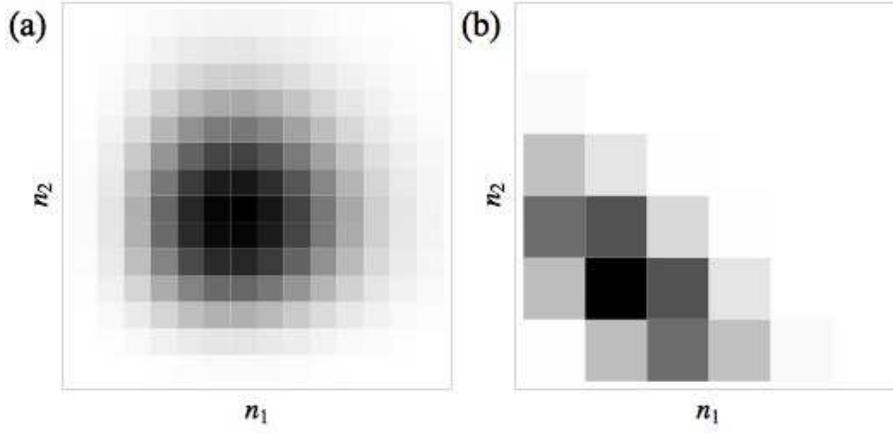}
\par\end{centering}
\caption{Two-photon-number probability density for (a) the harmonic oscillator
case with $z=3.5$ and (b) the case $\mu=5$ with $z=10^{5}.$}
\end{figure}

The linear entropy for our coherent state on a beamsplitter will be
(see \cite{Hoffmann2018c})
\begin{equation}
S(|z|)=1-\sum_{r_{1}=0}^{\infty}\sum_{r_{2}=0}^{\infty}|M(r_{1},r_{2};z)|^{2}.\label{eq:3.16}
\end{equation}
with
\begin{equation}
M(r_{1},r_{2};z)=\sum_{\kappa=0}^{\infty}G(\kappa+r_{1},r_{1};z)G^{*}(\kappa+r_{2},r_{2};z).\label{eq:3.17}
\end{equation}
and
\begin{align}
G(k,r;z) & =\alpha_{k}^{(5)}(z)\frac{1}{2^{\frac{k}{2}}}\binom{k}{r}^{\frac{1}{2}}=\frac{1}{\sqrt{F^{(5)}(z)}}\frac{z^{k}}{D_{k}^{(5)}}\frac{1}{2^{\frac{k}{2}}}\binom{k}{r}^{\frac{1}{2}}.\label{eq:3.18}
\end{align}
We take the sums to $\kappa=10$ and $r_{1}=r_{2}=10.$

We plot the linear entropy to $|z|=10^{5}$ in Figure 9. We see that
it remains small, indicating a low degree of entanglement.

\begin{figure}
\noindent \begin{centering}
\includegraphics[width=8cm]{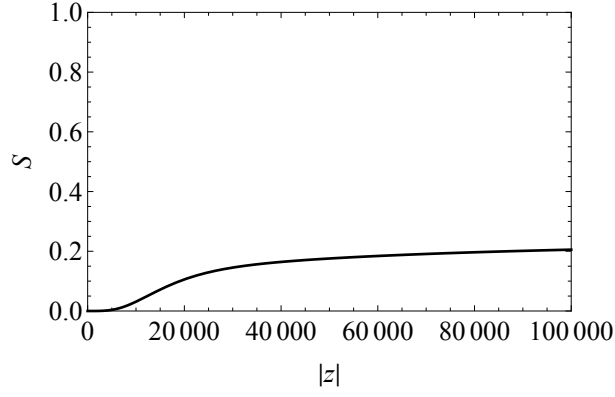}
\par\end{centering}
\caption{Linear entropy for $\mu=5$ on a beamsplitter.}
\end{figure}

\subsection{Heisenberg uncertainty principle}

We use results from \cite{Hoffmann2018c} (with $\alpha_{k}=\alpha_{k}^{(5)}(z)$)
\begin{eqnarray}
\langle\,z\,|\,\hat{x}\,|\,z\,\rangle & = & \sum_{k_{1}=0}^{\infty}\sum_{k_{2}=0}^{\infty}\alpha_{k_{1}}^{*}M_{k_{1}k_{2}}^{(x)}\alpha_{k_{2}},\nonumber \\
\langle\,z\,|\,\hat{x}^{2}\,|\,z\,\rangle & = & \sum_{k_{1}=0}^{\infty}\sum_{k_{2}=0}^{\infty}\alpha_{k_{1}}^{*}M_{k_{1}k_{2}}^{(x^{2})}\alpha_{k_{2}},\nonumber \\
\langle\,z\,|\,\hat{p}\,|\,z\,\rangle & = & \sum_{k_{1}=0}^{\infty}\sum_{k_{2}=0}^{\infty}\alpha_{k_{1}}^{*}M_{k_{1}k_{2}}^{(p)}\alpha_{k_{2}},\nonumber \\
\langle\,z\,|\,\hat{p}^{2}\,|\,z\,\rangle & = & \sum_{k_{1}=0}^{\infty}\sum_{k_{2}=0}^{\infty}\alpha_{k_{1}}^{*}M_{k_{1}k_{2}}^{(p^{2})}\alpha_{k_{2}},\label{eq:3.19}
\end{eqnarray}
with
\begin{eqnarray}
M_{k_{1}k_{2}}^{(x)} & = & \int_{0}^{\infty}dx\,\langle\,5+6k_{1}\,|\,x\,\rangle x\langle\,x\,|\,5+6k_{2}\,\rangle,\nonumber \\
M_{k_{1}k_{2}}^{(x^{2})} & = & \int_{0}^{\infty}dx\,\langle\,5+6k_{1}\,|\,x\,\rangle x^{2}\langle\,x\,|\,5+6k_{2}\,\rangle,\nonumber \\
M_{k_{1}k_{2}}^{(p)} & = & \int_{0}^{\infty}dx\,\langle\,5+6k_{1}\,|\,x\,\rangle(-i\frac{d}{dx})\langle\,x\,|\,5+6k_{2}\,\rangle,\nonumber \\
M_{k_{1}k_{2}}^{(p^{2})} & = & \int_{0}^{\infty}dx\,\langle\,5+6k_{1}\,|\,x\,\rangle(-\frac{d^{2}}{dx^{2}})\langle\,x\,|\,5+6k_{2}\,\rangle,\label{eq:3.20}
\end{eqnarray}
all integrated only on the right half-plane. With the range of $|z|$
considered, we may cut off the sums at $k_{1}=k_{2}=6$ with negligible
error. Once these expectations are calculated, we use
\begin{align}
\sigma_{x} & =\sqrt{\langle\,x^{2}\,\rangle-\langle\,x\,\rangle^{2}},\nonumber \\
\sigma_{p} & =\sqrt{\langle\,p^{2}\,\rangle-\langle\,p\,\rangle^{2}}.\label{eq:3.21}
\end{align}

The functions $\sigma_{x}(z)$ and $\sigma_{p}(z)$ and their product
are plotted in Figure 10 for real values of $z.$ We see no squeezing
in this regime, as the standard deviations remain larger than $1/\sqrt{2}$.
The product $\sigma_{x}\sigma_{p}$ remains greater than $1/2.$

\begin{figure}
\noindent \begin{centering}
\includegraphics[width=12cm]{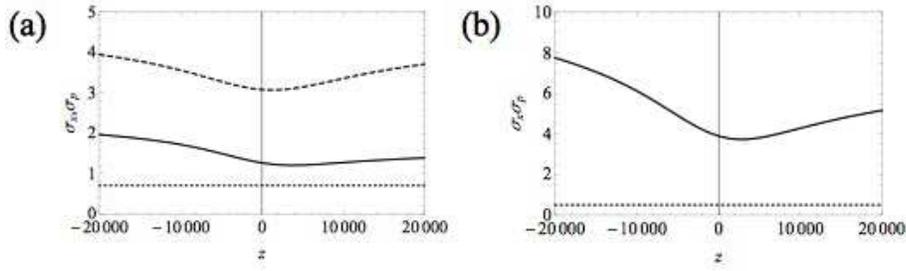}
\par\end{centering}
\caption{The spreads $\sigma_{x}(z)$ (solid) and $\sigma_{p}(z)$ (dashed)
are plotted in (a), compared to $1/\sqrt{2}$ (dotted) for real $z$.
Their product is plotted in (b) for real $z$ (solid) compared to
$1/2$ (dotted).}
\end{figure}

\subsection{Number statistics: Mandel $Q$ parameter}

As in \cite{Hoffmann2018c}, we define a number of excitations operator
by
\begin{equation}
N\,|\,5+6k\,\rangle=k\,|\,5+6k\,\rangle.\label{eq:3.22}
\end{equation}
The Mandel $Q$ parameter, defined by
\begin{equation}
Q=\frac{\Delta N^{2}-\langle\,N\,\rangle}{\langle\,N\,\rangle}=\frac{\langle\,N^{2}\,\rangle-\langle\,N\,\rangle^{2}-\langle\,N\,\rangle}{\langle\,N\,\rangle},\label{eq:3.23}
\end{equation}
is zero for Poissonian statistics, positive for super-Poissonian statistics
and negative for sub-Poissonian statistics.

We calculate the expectations by
\begin{align}
\langle\,N\,\rangle & =\sum_{k=0}^{\infty}|\alpha_{k}^{(5)}(z)|^{2}\,k,\nonumber \\
\langle\,N^{2}\,\rangle & =\sum_{k=0}^{\infty}|\alpha_{k}^{(5)}(z)|^{2}\,k^{2}.\label{eq:3.24}
\end{align}
The results are plotted in Figure 11, showing Poissonian statistics
only for $z=0$ and sub-Poissonian statistics for greater values of
$|z|.$

\begin{figure}
\noindent \begin{centering}
\includegraphics[width=8cm]{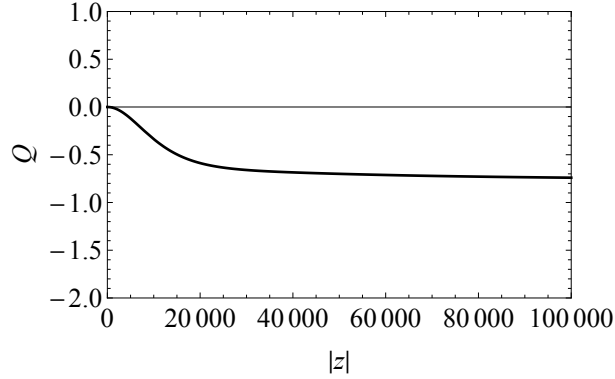}
\par\end{centering}
\caption{Mandel $Q$ parameter for $\mu=5.$}
\end{figure}

\section{Comparison with other work}

Fernández C. \textit{et al}. \cite{Fernandez2018b} considered the
same 4-step construction, with the choices in Eq. (\ref{eq:2.22}),
that we have used in this paper. They constructed coherent states
for linear ladder operators defined by
\begin{equation}
\mathcal{L}^{-}\,|\,\nu\,\rangle=\sqrt{\nu-1}\,|\,\nu-2\,\rangle,\quad\mathcal{L}^{+}\,|\,\nu\,\rangle=\sqrt{\nu+1}\,|\,\nu+2\,\rangle,\quad\mathrm{for}\ \nu=1,3,\dots,\label{eq:4.1}
\end{equation}
when acting only on the part of the spectrum isospectral with the
truncated harmonic oscillator. Since they lower or raise the index
by 2, they can be considered linearized versions of our ladder operators
$\tilde{L}$ and $\tilde{L}^{\dagger},$ respectively. These authors
use the displacement operator to define their coherent states, but
since the structure is just that of the harmonic oscillator, this
gives the same result as the Barut-Girardello definition \cite{Barut1971}.
Of course it is the fact that the position wavefunctions differ from
those of the harmonic oscillator that leads to new and interesting
results.

These authors find squeezing in $p,$ whereas we found no squeezing
for real $z.$ For their coherent state on one arm of a beamsplitter,
they found a linear entropy that was largely flat as a function of
$|z|,$ with values close to $0.5,$ whereas we found a function at
first decreasing, then flat, close to 0.2 asymptotically. Thus the
degree of entanglement was low for both cases. Our variation was over
a much larger scale of $|z|,$ understood from the presence of large
denominators, as explained for energy expectations in Section III
A.

Two coherent states of completely different construction, as we have
here, are not expected to have similar physical properties at equal
$z$ values. Instead, the energy expectation allows a more meaningful
comparison. The energy expectation for the coherent states constructed
by Fernández C. \textit{et al}. \cite{Fernandez2018b} would be
\[
E_{F}(|z|)=3+2|z|^{2}.
\]
They consider $|z|$ values as large as 3, giving a maximum energy
of 21. This, from Figure 3, is comparable to the energies we were
considering.

We did not expect close agreement with Fernández C. \textit{et al}.
\cite{Fernandez2018b}, as the structure of the coherent states differed
(our ladder operators change the index by 6 rather than 2) and the
coefficients in the superpositions are distinctly different.

\section{Conclusions}

We constructed the coherent states that are eigenvectors of the $\tilde{C}$
ladder operator with complex eigenvalue $z$ and lowest weight $\mu=5.$

The energy expectation was found to vary slowly with $|z|.$ Thus
we chose to consider coherent states for the other tests with $z$
values sufficiently large to make the energy expectation sufficiently
different from the ground state energy. We plotted the time-dependent
position probability density over one period for $z=10^{5}$ and found
a great deal of structure. Such was also the case for the even and
odd cat state densities.

The Wigner function for $z=500$ displayed a great deal of structure
and showed no definitive regions of negative value. So we could not
claim to have seen the signature of nonclassical behaviour.

For the coherent states on one arm of a beamsplitter, the two ``photon''
probability distribution was seen to not factorize and was plotted
for $z=10^{5}.$ The linear entropy was found to be significantly
low compared to unity up to $z=10^{5},$ indicating a low degree of
entanglement.

Calculation of the standard deviations in position and momentum for
real $z$ with $|z|\leq20\,000$ showed no squeezing. The Heisenberg
uncertainty principle was satisfied as a check on our calculations.

Calculation of the Mandel $Q$ parameter showed the number statistics
of our coherent states to be sub-Poissonian up to $z=10^{5},$ Poissonian
only for $z=0.$

We conclude that the majority of these indicators suggested non-classical
behaviour for these coherent states.

We compared our results with those of Fernández C. \textit{et al}.
\cite{Fernandez2018b} for the same basis states and position wavefunctions
but quite different coherent states. As expected, our results differed
from theirs.
\begin{acknowledgments}
IM was supported by Australian Research Council Discovery Project
DP 160101376. YZZ was supported by National Natural Science Foundation
of China (Grant No. 11775177). VH acknowledges the support of research
grants from NSERC of Canada. SH receives financial support from a
UQ Research Scholarship.
\end{acknowledgments}

\bibliographystyle{vancouver}

\end{document}